\documentstyle[preprint,revtex]{aps}
\begin{document}
\draft
\preprint{IPT-EPFL preprint}
\begin{title}
A Remark On Interacting Anyons
In Magnetic Field
\end{title}
\author{D. F. Wang and C. Gruber}
\begin{instit}
Institut de Physique Th\'eorique\\
\'Ecole Polytechnique F\'ed\'erale de Lausanne\\
PHB-Ecublens, CH-1015 Lausanne-Switzerland.
\end{instit}
\begin{abstract}
In this remark, we note that the anyons, interacting with each other
through pairwise potential $\sum_{i\ne j} V(|\vec r_i - \vec r_j|)$ in
external magnetic field, exhibits a quantum group
symmetry $U_q(sl(2))$.
\end{abstract}

\narrowtext
There have been considerable interests in the quantum mechanics of
anyons in two dimensions. Anyons have been suggested to
be relevant to various interesting systems, such as the
quasiparticles in the Fractional Quantum Hall fluids,
the solitons in the nonlinear
sigma model\cite{arovas,wu,wu2,chen,jackiw,lei,chou,chin,mag,can}.
Due to the nontrivial
statistical interaction between the anyons, elementary
quantum mechanics of free anyon gas is far from being
understood fully\cite{wu,wu2}. For free anyons, the system is found to exhibit
the conformal symmetry\cite{jackiw}. Attempts have been made to find the energy
spectrum of the free anyon gas, however,
only in the case of two anyons, the complete solution
is available by far\cite{wu,wu2}. In presence of external magnetic
field, the ladder operator approach can provide us with
some portion of the energy spectrum and the wavefunctions
of the anyon gas\cite{mag}.
In this letter, we present a simple quantum group symmetry for
interacting anyons in external magnetic field.

The anyons in external magnetic field are described
in terms of the following Hamiltonian:
\begin{equation}
H=\sum_{i=1}^N {1\over 2} [\vec p_i - \alpha
\sum_{j(\ne i)=1}^N { \vec k \times \vec r_{ij} \over |\vec r_{ij}|^2}-
{e B \over 2}  \vec k \times \vec r_i]^2,
\end{equation}
where we assume that $N$ anyons move in a two dimensional infinite
plane, and the statistical parameter $\alpha$ is within the range $[-1,1]$.
For the constant external magnetic field $\vec B=B\vec k$, we choose
the symmetric gauge, such that the vector field
is $\vec A_i = {B\over 2} \vec k \times \vec r_i$.
For these anyons, the bosonic wavefunctions are used as their base functions,
\begin{equation}
\phi (\cdots,\vec r_i,\cdots,\vec r_j,\cdots)
= \phi (\cdots,\vec r_j,\cdots,\vec r_i,\cdots),
\end{equation}
and the anyons are composite particles of the bosons attached with flux.
The anyon spectrum is determined by the following
eigenequation:
\begin{equation}
H \phi(\vec r_1, \vec r_2, \cdots, \vec r_N) = E \phi(\vec r_1,\vec r_2,
\cdots,\vec r_N).
\end{equation}

For the anyons, one may perform the following well-known singular statistical
gauge transformation $U=e^{-i\alpha \sum_{k< l} \theta_{kl}}$:
\begin{equation}
\phi'(\vec r_1,\vec r_2,\cdots)= e^{i\alpha \sum_{k<l}\theta_{kl}}
\phi(\vec r_1,\vec r_2, \cdots),
\label{eq:anyonfunction}
\end{equation}
where $\theta_{ij}$ is the angle between
the vector $\vec r_{ij}$ and the x-axis,
$\vec r_{ij} = \vec r_i -\vec r_j = r_{ij}(\cos\theta_{ij} \vec x
+ \sin\theta_{ij} \vec y)$.
The eigenenergy equation can be translated into
\begin{equation}
H' \phi' = E \phi',
\end{equation}
with
\begin{equation}
H'=U^\dagger H U = \sum_{i=1}^N {1\over 2}
(\vec p_i -{e B\over 2} \vec k \times \vec r_i)^2.
\end{equation}
For the multivalued wavefunction $\phi'$, additional phase
factor shows up when one interchanges two patricles,
\begin{equation}
\phi'(\cdots,x_i,\cdots,x_j,\cdots)=e^{\pm i\alpha\pi}\phi'(
\cdots,x_j,\cdots,x_i,\cdots).
\end{equation}
One can see that the wavefunction $\phi'$ is invariant
under the change $\alpha\rightarrow \alpha+2$.
In particular, when the statistical parameter $\alpha =1$,
the wavefunction $\phi'$ is antisymmetric, describing
$N$ fermions in external magnetic field. For $N$ fermions,
the energy spectrum of the system is given by the sum
of single particle energies, as long as the Pauli exclusion
principle is properly taken into account when one
constructs the Slater determinant. For one electron in
external magnetic field, we have the well-known Landau
problem in two dimenions\cite{land}, which is also the problem of
Integer Quantum Hall Effect when the magentic field $B$ is
so strong, compared to the electron-electron Coulomb interaction
and the disorder in real two dimensional electron gas,
with small ratio of electron-electron interaction to disorder.

In the Landau problem,
the system is not translational invariant.
However, it has been known for a long time
that the system is invariant under the
magnetic translation group. Considering the Hofstadter
problem ( a free electron
hopping on two dimensional lattice in constant magnetic field),
Wiegmann and Zabrodin show that the magnetic
translations can be constructed with the generators
of the quantum group $U_q(sl(2))$\cite{wie}. Applying
the representation, they obtain the Bethe-Ansatz.
Quite recently, similar discussions have been carried
out for the Landau problem, where the quantum group
symmetry $U_q(sl(2))$, the $W$ algebra and the area-presevering
differomorphism were studied\cite{kogan}.

When one has arbitrary statistical parameter $-1< \alpha <1$,
the system consists of particles of intermediate statistics.
The anyons are interacting with each other through nontrivial
statistical gauge term, and the eigenenergy problem can no longer be
reduced to a single particle problem.
However, in the following,
we would like to demonstrate that the system of the
anyons still exhibits a magnetic translation group,
from which the quantum group symmetry $U_q(sl(2))$
can be constructed. In fact, one will see that
the quantum group symmetry will survive, even when the anyons
are interacting with each other through
pairwise potential $\sum_{i\ne j} V(r_{ij})$.

In presence of pairwise interaction between the anyons,
the Hamiltonian for the
anyons takes the following form:
\begin{equation}
H_N=\sum_{i=1}^N {1\over 2} [\vec p_i - \alpha
\sum_{j(\ne i)=1}^N { \vec k \times \vec r_{ij} \over |\vec r_{ij}|^2}-
{e B \over 2}  \vec k \times \vec r_i]^2 + \sum_{i\ne j} V(r_{ij}).
\label{eq:anyon}
\end{equation}
With the singular statistical gauge transformation $U$, the Hamiltonian
becomes
\begin{equation}
H_N'=U^\dagger H_N U
=\sum_{i=1}^N {1\over 2} (\vec p_i-{eB\over 2} \vec k \times
\vec r_i)^2 + \sum_{i\ne j} V(r_{ij}).
\label{eq:complete}
\end{equation}
Let us construct following generators of magnetic translation:
\begin{equation}
T_{\vec \zeta}= \prod_{k=1}^N \exp[{ \zeta_1(\partial_{x_k}+
eBiy_k/2)  + \zeta_2 (\partial_{y_k}
- eBix_k/2)}],
\label{eq:magnetic}
\end{equation}
where $\vec \zeta = \zeta_1 \vec x + \zeta_2 \vec y $ is a two
dimensional vector, and the symmetric gauge
for the vector field is used, $\vec A_i = {B\over 2} \vec k
\times \vec r_i$. The generator $T_{\vec \zeta}$ is the product of
the generators of magnetic translation of each single particle.
They satisfy the following group property\cite{zak}:
\begin{equation}
T_{\vec \zeta} T_{\vec \eta} = \exp[ {-iN {\vec B\over 2} \cdot (
\vec \zeta \times \vec \eta)}] ~ T_{\vec \zeta + \vec \eta}.
\end{equation}
The generators commute with the pairwise potential term,
\begin{equation}
[T_{\vec \zeta}, \sum_{i\ne j} V(|\vec r_{ij}|)]=0.
\end{equation}
One can verify these generators commute with the Hamiltonian $H_N'$ given
by Eq.~(\ref{eq:complete}),
\begin{equation}
[H_N', T_{\vec \zeta}]=0.
\end{equation}
Moreover, with the simple relation
$[\sum_{i=1}^N \partial_{x_i}, \sum_{k<l} \theta_{kl}]=0$
and $[\sum_{i=1}^N \partial_{y_i},\sum_{k<l}\theta_{kl}]=0$,
one can check that the singular statistical gauge
transformation $U$ commutes with the generators of
magnetic translation group,
\begin{equation}
[T_{\vec \zeta}, U]=0.
\end{equation}
With these equations, we therefore conclude that
the magnetic translation group generators also commute
with the original anyon Hamiltonian Eq.(\ref{eq:anyon})
\begin{equation}
[H_N, T_{\vec \zeta}]=0,
\end{equation}
i.e., the anyon gas is invariant under the magnetic translation.

With the generators of magnetic translation group,
one can construct the generators of the quantum group
$U_q(sl(2))$ as follows\cite{wie}:
\begin{eqnarray}
&&J_+={1\over (q-q^{-1}) } (\alpha T_{\vec a} + \beta T_{\vec b}),\nonumber\\
&&J_{-}={1\over (q-q^{-1})}
(\gamma T_{-\vec a} + \delta T_{-\vec b}),\nonumber\\
&&q^{2J_3}=T_{\vec b -\vec a},\nonumber\\
&&q^{-2J_3} = T_{\vec a-\vec b},
\end{eqnarray}
where $q=\exp[iN {\vec B \over 2} \cdot (\vec a \times \vec b)]$,
$\alpha\delta=\beta \gamma=-1$, and $\vec a, \vec b$ are two vectors
defined in the two dimensional plane.
Using the group property of the magnetic translation generators,
one can check that the following commutation relations of the
$U_q(sl(2))$\cite{ge,fa,sk} quantum group hold
\begin{eqnarray}
&&q^{J_3} J_{\pm} q^{-J_3} =q^{\pm 1} J_{\pm}\nonumber\\
&&[J_+,J_{-}]={ q^{2J_3} - q^{-2J_3} \over q-q^{-1} }.
\end{eqnarray}
The generators for the quantum group $U_q(sl(2))$ also commute with the
original anyon Hamiltonian Eq.~(\ref{eq:anyon}), and we therefore
conclude that the interacting anyons in magnetic field also exhibits
a quantum group $U_q(sl(2))$,
\begin{eqnarray}
&&[H_N, J_{\pm}]=0,\nonumber\\
&&[H_N, J_3]=0.
\end{eqnarray}
The above two equations are the main results
we would like to make remarks of in this letter.
Finally, we note that the quantum group symmetry is also valid
for the interacting electron system in strong magnetic field.
The many particle system is the problem of Fractional Quantum
Hall when the ratio of disorder to the electron-electron interaction
is small. For the interacting electron system, the Hamiltonian
is given by
\begin{equation}
H=\sum_{i=1}^N {1\over 2} (\vec p_i -e {\vec B\over 2} \times \vec r_i)^2
+ \sum_{i\ne j} {e^2\over r_{ij}},
\end{equation}
and the wavefunction $\phi'$ is totally antisymmetric when interchanging
any pair of particles, which corresponds to $\alpha=1$.

In summary,
for arbitrary intermediate
statistics, the interacting anyon gas in external magnetic
field is invariant under one magnetic translation group,
as the singular statistical gauge transformation commutes
with such magnetic translations.
The interacting anyons hence
exhibits a quantum group symmetry $U_q(sl(2))$, which
can be readily constructed from the magnetic
translation group.

We would like to thank the Swiss National Science Foundation
for the financial support.
We also would like to thank Ian I. Kogan for communication.

\end{document}